\documentclass[showpacs,preprintnumbers,amsmath,amssymb,twocolumn]{revtex4}
\usepackage{graphicx}
\usepackage{dcolumn}
\usepackage{bm}

\begin{document}

\title{Fluorescence Intermittency of A Single Quantum System and
Anderson Localization}

\author{Xiang Xia}
\author{Robert J. Silbey}
  \email{silbey@mit.edu}
\affiliation{Department of Chemistry, Massachusetts Institute of
Technology, Cambridge, MA
02139}%

\begin{abstract}
The nature of fluorescence intermittency for semiconductor quantum
dots (QD) and single molecules (SM) is discussed from the
viewpoint of Anderson localization. The power law distribution for
the \emph{on} time is explained as due to the interaction between
QD/SM with a random environment. In particular, we find that the
\emph{on}-time probability distribution behaves differently in the
localized and delocalized regimes. They, when properly scaled, are
\emph{universal} for different QD/SM systems. The \emph{on}-time
probability distribution function in the delocalized QD/SM regime
can be approximated by power laws with exponents covering $-2\le m
<0$.
\end{abstract}
\maketitle

Recent development in nano-fabrication and measurement has made it
possible to probe directly the dynamics of a single quantum system
and its coupling with local environment. For optical measurements,
in particular, fluorescence intermittency (FI) occurs in a wide
class of single quantum systems including semiconductor
nanocrystals \cite{KFHGN,SNLEWB,VOO,CMB,IBC} and single molecules
\cite{SCB1,SCB2,SCB3}. This optical phenomenon is experimentally
found to be a robust and fundamental property of single photon
emitters. The random switching between an emitting (\emph{on}) and
a non-emitting (\emph{off}) state is characterized by
\emph{on}-time and \emph{off}-time distribution functions.
Surprisingly, the distribution functions can often be fitted by
power laws instead of exponentials. The exponents of the power
laws vary from system to system and they are almost independent of
temperature. To explain this unusual behavior, Marcus \emph{et.
al.} \cite{TM1,FM} recently proposed a model based on diffusion in
energy space. An exponent of $-1.5$ is readily obtained for
one-dimensional diffusion processes. Deviation from $-1.5$ is
explained by anomalous diffusion, which has its origin from the
Cole-Davison dielectric medium $\beta\ne1$. However, this scenario
would predict exponents that are temperature dependent. Recent
experiments also found that exponents $\sim-2.0$ for the \emph{on}
time exist in many different single emitters
\cite{KFHGN,CMB,SCB1,SCB2,SCB3}, which are imbedded in various
media. The temperature-independence of power laws with
$m_{\text{on/off}}\ne -1.5$ and robustness of $m_{\text{on}}=-2.0$
need to be explained.

In this work, we provide a microscopic Hamiltonian model which
does not build on the assumption of anomalous diffusion in energy
space and the quadratic form of diffusion potential. The purpose
of this work is to lay the foundation for a quantum-mechanical
theory of FI. In a classic paper \cite{A}, Anderson considered a
transport process which involves solely quantum-mechanical motion
in a random lattice. His quantum-mechanical treatment of the
process leads to the concept of Anderson localization, which has
been applied extensively to the problem of transport in a random
environment. Here, we view the blinking as a charge hopping
process between the QD/SM and the traps in the environment. We
find that the FI can be explained in the framework of the Anderson
localization. The basic idea is the following: due to the coupling
to a random environment, the on-site energy of a QD/SM is
renormalized and becomes a random variable. In addition, random
configurations of the local environment lead to a distribution of
decaying (hopping) rates for the QD/SM. This is the physical
origin of the \emph{on}-time distribution.

Based on experimental results, we study the following simple
tight-binding Hamiltonian:
\begin{equation}
\hat{H}=\varepsilon_{d}\hat{d}^{\dagger}\hat{d}+\sum_{j}\varepsilon_{j}\hat{c}_{j}^{\dagger}\hat{c}_{j}
\label{eq:Hamiltonian}+\sum_{j}v_{j}\left(\hat{d}^{\dagger}\hat{c}_{j}+\hat{c}^{\dagger}_{j}\hat{d}\right)
\end{equation}
The ground state is defined such that the QD/SM is neutral. For a
QD this corresponds to a filled valence band. When an electron is
excited by photon to an excited state, it relaxes to the lowest
excited state on a time scale from hundreds of femtoseconds to
picoseconds. This lowest excited state is responsible for photon
emission and hopping to traps in the matrix. This state could be
$1S$ state for QDs and it may involve some surface states as well
\cite{KSMLB,KMMLB}. The excited electron completes a radiative
cycle on a time scale of the order inverse Rabi frequency:
$\Omega^{-1}_{\text{Rabi}}$. This time scale is much smaller than
the typical bin size, $\ge$0.2 ms, used in the experiments. Since
the excited electronic state is near resonance with the ground
state plus a photon, we define the \emph{on}-state as the
following: it is a dressed electronic state \cite{NOTE1} which has
the same bare energy as the excited electronic state,
$\varepsilon_{d}$, but with an infinite lifetime when it is
decoupled from the trap states. Physically, this means that when
there is no coupling to the environment, the excited electron
localizes inside the QD/SM and the QD/SM, being in the
\emph{on}-state, keeps emitting photons. Such an \emph{on}-state
is created by operator $\hat{d}^{\dagger}$. The bare on-site
energy $\varepsilon_{d}$ is renormalized due to interaction with
trap sites $j$. The coupling constant is $v_{j}$. The trap site
$j$ is created by operator $\hat{c}^\dagger_{j}$ and has a random
energy $\varepsilon_{j}$. Here we focus on the class of random
environments that are formed by topologically random network of
chemical bonds. Randomness could arise from a spatially
fluctuating potential due to charged impurities and coupling with
phonons that arise from deformation of a random lattice. The
details of randomness are irrelevant to the dynamical response of
the system as is known from scaling theory of localization. For
simplicity, we assume all traps follow the same energy
distribution function $p(\varepsilon)$.

Mathematically, the above definitions are equivalent to defining
the zeroth order Green's function of the \emph{on}-state as:
$G^{(0)}_{d}(\omega)=\frac{1}{\omega-\omega_{d}+i\eta}$, with
$\omega_{d}=\varepsilon_{d}/\hbar$, and $\eta$ being an
infinitesimally small and positive constant. Similarly, the zeroth
order Green's functions of the traps are
$G^{(0)}_{j}(\omega)=\frac{1}{\omega-\omega_{j}+i\eta}$. The decay
rate of the \emph{on}-state can be calculated from the full
Green's function $
G_{d}(\omega)=\frac{1}{\omega-\omega_{d}-\Sigma_{s}(\omega)}$
where the only self-energy of the model is
\begin{equation}
\Sigma_{s}(\omega)=\frac{1}{\hbar^{2}}\sum_{j}\frac{v_{j}^{2}}{\omega-\omega_{j}+i\eta}
\label{eq:Selfenergy}
\end{equation}
One obtains the renormalized energy and decay rate from the pole
equation. The results to $O(v_{j}^{2})$ are:
\begin{subequations}
\begin{equation}
E_{d}=\varepsilon_{d}+\sum_{j}\frac{v_{j}^{2}(\varepsilon_{d}-\varepsilon_{j})}{(\varepsilon_{d}-\varepsilon_{j})^{2}
+\eta^{2}} \label{eq:RenomalizedEnergy}
\end{equation}
\begin{equation}
\Gamma=\sum_{j}\frac{2v_{j}^{2}}{\hbar}\frac{\eta}{(\varepsilon_{d}-\varepsilon_{j})^{2}+\eta^{2}}
\label{eq:Decayrate}
\end{equation}
\end{subequations}
Since each $\varepsilon_{j}$ is a random variable, both the
renormalized energy $E_{d}$ and decay rate $\Gamma$ are random
variables. Physically, this corresponds to the situation in which
environment stays at a certain configuration for a period of time
during which a physical decay rate can be defined. The ensemble of
configurations that the environment takes on gives rise to a
distribution of decay rates. This picture is valid when static
disorder dominates or when the environment is ergodic.

To find the decay rate distribution function, $f(\Gamma)$, from
$p(\varepsilon_{j})$, one must perform a summation over random
variables $\varepsilon_{j}$. In addition, due to lattice
deformation, a random spatial distribution of traps must also be
assumed. For simplicity, we assume that all the traps follow the
same distribution function $\rho(\textbf{r})$. We will consider
two physical regimes, namely the delocalized and the localized
regimes, for the excited electron on the QD/SM according to the
framework of Anderson \cite{A}.

The delocalized regime corresponds to most of the experimentally
studied situations. Essentially, electrons excited inside the
QD/SM can make real transitions to the trap sites in the matrix
through the mechanism of quantum tunneling, and then become
localized. When this happens, the QD/SM becomes ionized. The
results in this regime, which will be shown later, support the
random resonance picture: on average, the energy of a QD/SM, is
far off-resonance with trap energies, i.e.
$\varepsilon_{d}\gg\varepsilon_{j}$. The electrons of QD/SM remain
localized (\emph{on} state) until a random fluctuation of
$\varepsilon_{j}$ makes it on-resonance with the QD/SM. This
random resonance could be realized through interaction with
phonons \cite{MA}. Since there is finite probability that
$\varepsilon_{d}-\varepsilon_{j}\sim\eta\sim0^{+}$, one cannot
perform the summations in the same way as in \cite{A}. Instead,
nontrivial summations of the whole expression (\ref{eq:Decayrate})
are required. The distribution function $f(\Gamma)$ can be
calculated by method of Fourier transformation:
\begin{equation}
f(\Gamma)=\int_{-\infty}^{\infty}\frac{dt}{2\pi}~\psi(t)~e^{-i\Gamma
t} \label{eq:Fgamma}
\end{equation}
The Fourier transformed probability distribution function
$\psi(t)$ is given by
\begin{eqnarray}
\psi(t)&=&\int_{-\infty}^{\infty}\prod_{j=1}^{N}
d\varepsilon_{j}p(\varepsilon_{j})\exp\left[it\sum_{k=1}^{N}\frac{2v_{k}^{2}}{\hbar}
\frac{\eta}{(\varepsilon_{d}-\varepsilon_{k})^{2}+\eta^{2}}\right]\nonumber\\
&=&\exp\left\{ N\ln\mathcal{I}\right\} \label{eq:Psi}
\end{eqnarray}
where $N$ is the number of traps and
\begin{equation}
\mathcal{I}=\int_{-\infty}^{\infty}d\varepsilon\int
d\textbf{r}~p(\varepsilon)\rho(\textbf{r})\exp\left[i\frac{2tv(\textbf{r})^{2}}{\hbar}
\frac{\eta}{(\varepsilon_{d}-\varepsilon)^{2}+\eta^{2}}\right]\nonumber
\label{eq:Integral}
\end{equation}
The integral $\mathcal{I}$ is obtained under the assumptions made
above, i.e. identical energy and spatial distributions of traps.
This integral needs to be done carefully so that we can express it
in powers of $\eta$. For a real physical system coupled to a
disordered environment, $\eta$ is small but nonetheless not
strictly zero. Physically, this is due to competing non-radiative
relaxation channels which contribute to a small but finite width
of resonance. The value of $\eta$ can be calculated in a more
sophisticated model, however, in our model we take it as a fitting
parameter. Changing variable to the dimensionless
$u=\frac{\hbar}{2\pi v(\textbf{r})^{2}|t|}\varepsilon$ and
noticing $\frac{1}{\pi}\frac{\eta}{(u-u_{d})^{2}+\eta^{2}}$ can be
approximated by an impulse function centered at $u=u_{d}$ as
$\eta$ is sufficiently small, where $u_{d}=\frac{\hbar}{2\pi
v(\textbf{r})^{2}|t|}\varepsilon_{d}$, we can separate the
integral into three parts: $(-\infty,u_{d}-\eta/2]$,
$(u_{d}-\eta/2,u_{d}+\eta/2)$ and $[u_{d}+\eta/2,+\infty)$ to
obtain the expansion:
\begin{widetext}
\begin{equation}
\mathcal{I}=1-\eta\left[1-\cos\left(\frac{1}{\eta}\right)-i~\text{sign}(t)\sin\left(\frac{1}{\eta}\right)\right]
p(\varepsilon_{d})\frac{2\pi}{\hbar}\langle
v(\textbf{r})^{2}\rangle |t|+O(\eta^{3})\nonumber
\end{equation}
\end{widetext}
Since we are interested in case where there are a large number of
traps $N\gg1$, by substituting the expansion to Eq.(\ref{eq:Psi})
we rewrite the unknown parameter $N\eta$ as $N\eta=a_{0}n$
\cite{A,NOTE2}, where $n$ is the density of traps and $a_{0}$ is a
fitting parameter. Absorbing the numerical constant in the square
bracket of $\mathcal{I}$ into $a_{0}$ and neglecting an
unimportant imaginary part as is justified in this delocalized
regime, we get
\begin{equation}
\psi(t)=\exp\left\{-\frac{2\pi}{\hbar}\langle
v(\textbf{r})^{2}\rangle p(\varepsilon_{d})na_{0}|t|\right\}
\end{equation}
The probability distribution function $f(\Gamma)$ in the
delocalized QD/SM regime:
\begin{equation}
f(\Gamma)=\frac{1}{\pi}\frac{\Gamma_{0}}{\Gamma^{2}+\Gamma_{0}^{2}}
\label{eq:fGamma2}
\end{equation}
is Lorentzian, with the characteristic decay rate $\Gamma_{0}$
given by:
\begin{equation}
\Gamma_{0}=\frac{2\pi}{\hbar}\langle v(\textbf{r})^{2}\rangle
p(\varepsilon_{d})na_{0} \label{eq:Gamma0}
\end{equation}
The \emph{on}-time distribution function can be calculated as
$P_{\text{on}}(t)=\int_{0}^{\infty}d\Gamma \exp(-\Gamma
t)f(\Gamma)$ which turns out to be:
\begin{equation}
P_{\text{on}}(t)=-\frac{2\Gamma_{0}}{\pi}\left[\cos(\Gamma_{0}t)\text{ci}(\Gamma_{0}t)
+\sin(\Gamma_{0}t)\text{si}(\Gamma_{0}t)\right] \label{eq:Pon2}
\end{equation}
where $\text{ci}$ and $\text{si}$ are cosine and sine integrals
\cite{GR} respectively.

\begin{figure}
\resizebox{3.2in}{2in}{\includegraphics{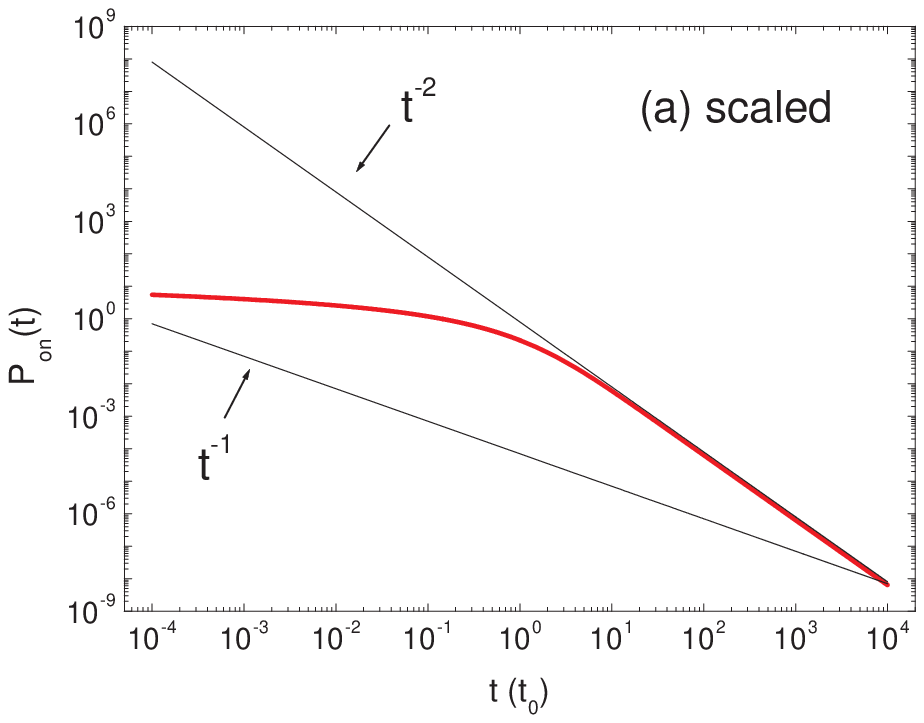}}
\resizebox{3.2in}{2in}{\includegraphics{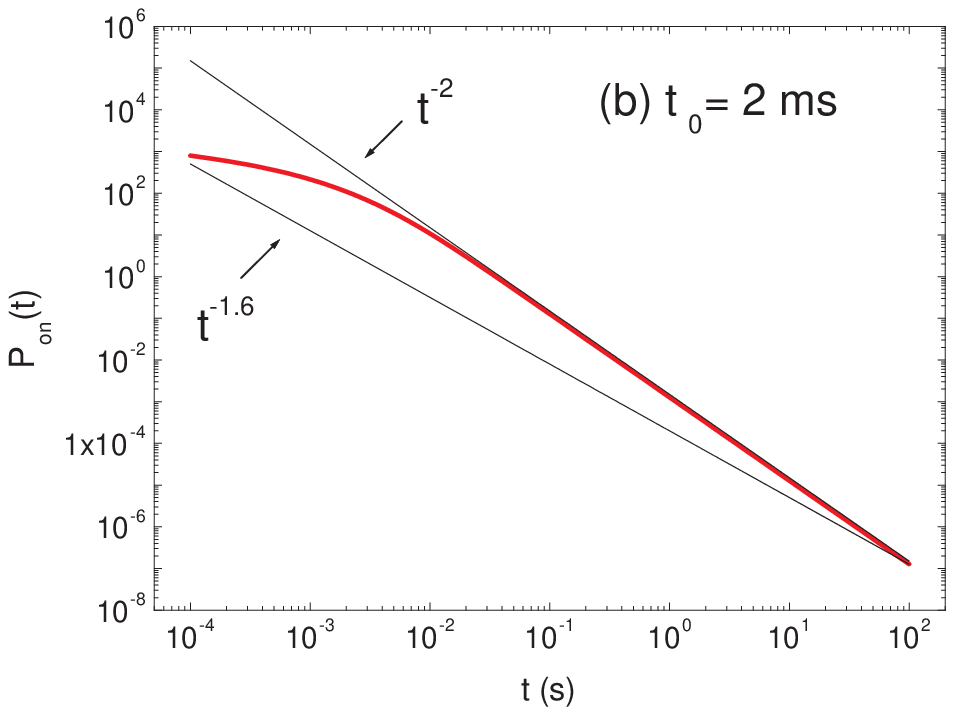}}
\resizebox{3.2in}{2in}{\includegraphics{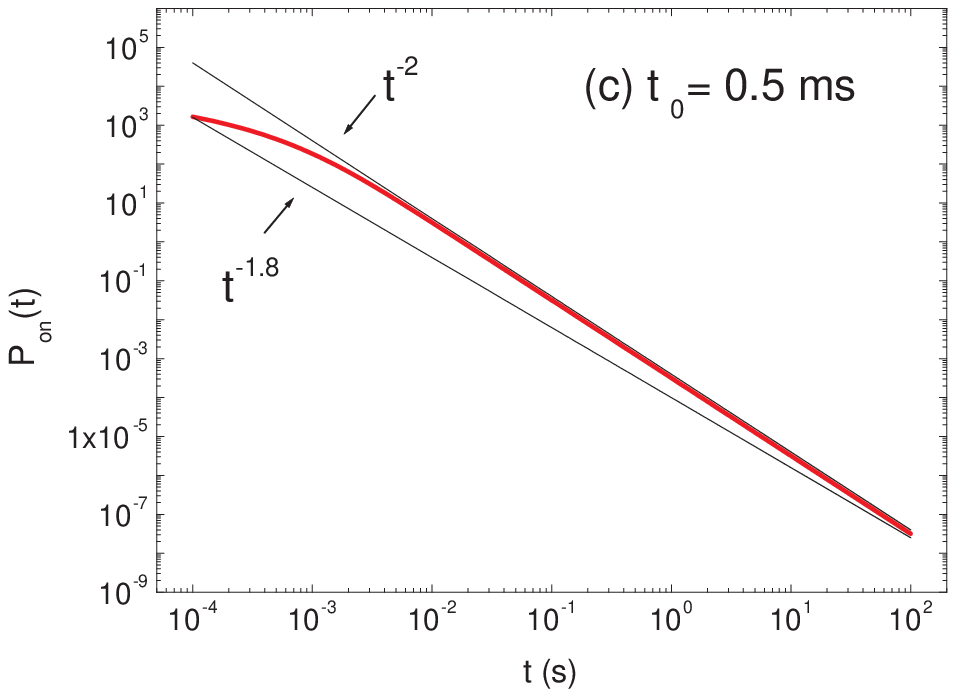}}
\caption{\label{fig:1} Probability distribution functions for the
\emph{on} time in the delocalized regime. (a) Universal
\emph{on}-time probability distribution function obtained by
scaling in a characteristic \emph{on}-state lifetime
$t_{0}=1/\Gamma_{0}$. (b) $P_{\text{on}}(t)$ for a QD/SM with a
lifetime $2\text{ms}$ and a fixed observation window. The
exponents of the fitting power laws lie between $-1.6$ and $-2$.
(c) $P_{\text{on}}(t)$ of a QD/SM with a shorter lifetime
$0.5\text{ms}$. The exponents of the fitting power laws lie
between $-1.8$ and $-2$.}
\end{figure}

The expression for $P_{\text{on}}(t)$ in (\ref{eq:Pon2}) has
several physical implications. First of all, it depends only on
one parameter $\Gamma_{0}$ (this is also true for
$P_{\text{on}}(t)$ in the localized regime (\ref{eq:Pon1}), which
only depends on the parameter $\gamma_{0}$). The characteristic
decay rate $\Gamma_{0}$, is purely due to quantum tunneling and
depends on the degree of disorder $p(\varepsilon_{d})\langle
v(\textbf{r})^{2}\rangle$. This is a generic feature of Anderson
localization theory. Only the energy fluctuation of traps
evaluated at the bare QD/SM on-site energy, $p(\varepsilon_{d})$,
comes into the final result. This supports the random resonance
picture discussed above. The one-parameter theory also implies
that experimental results for different QD/SM systems, when
plotted in unit of characteristic \emph{on}-time lifetime
$t_{0}=1/\Gamma_{0}$, follow a \emph{universal} distribution
Fig.\ref{fig:1}(a). In the long time limit, $t\gg t_{0}$,
$P_{\text{on}}(t)\sim t^{-2}$. The exponent of $-2$ comes from the
peak of the Lorentzian distribution for decay rates, which
physically indicates the resonant scattering condition. This $-2$
exponent is also robust and it does not depend on the properties
of matrix or emitters \cite{SCB1}. In experiments, the observation
time is fixed so that one probes only a certain range ($3$ to $4$
orders of magnitude in physical time) of the whole probability
distribution. The location of the window in the universal
distribution depends on $\Gamma_{0}$, which is \emph{not}
universal. If one fits the experimental data with power laws, then
the exponents of the power laws will implicitly depend on
$\Gamma_{0}$ except for the $m_{\text{on}}=-2$ case. From
(\ref{eq:Gamma0}) we see that $\Gamma_{0}$ depends on properties
of the QD/SM and its embedding matrix. This is one major
difference between our theory and the diffusion-controlled
electron transfer theory \cite{TM1} for the \emph{on}-time
statistics. In the latter, the exponents of the fitting power laws
come from properties of the dielectric medium alone, i.e. the
$\beta$ parameter of the Cole-Davison equation, which depends on
temperature. As $\Gamma_{0}$ rises with increasing coupling
strength between the QD/SM and its local environment, power law
exponents observed for SMs would in principle larger than (in
absolute value) that of well-coated QDs. This is indeed consistent
with current experimental results\cite{SNLEWB,SCB1}.

The quantum mechanical theory of FI allows a weak
temperature-dependence of the \emph{on}-time probability density.
This comes from the fact that the energy fluctuation of trap
sites, besides the mechanisms discussed above, has a temperature
component. This dependence is usually very weak since the typical
energy difference between $\varepsilon_{d}$ and $\varepsilon_{j}$
is on the order of $1\sim2$ eV which is much larger than the
magnitude of thermal broadening around $\varepsilon_{j}$.
Increasing temperature slightly increases $p(\varepsilon_{d})$
since $\varepsilon_{d}\gg\varepsilon_{j}$ and as a result
$\Gamma_{0}$ increases. So higher temperature would slightly
decrease $m_{\text{on}}$ towards $-2$ for the long time.

In the case that the excited QD/SM electron is localized, then we
predict very long \emph{on} times. This corresponds to the
off-resonance situation, $\varepsilon_{d}-\varepsilon_{j}\gg \eta,
\forall j$. One can calculate the probability distribution of
$\Gamma=\frac{2\eta}{\hbar}\sum_{j}\frac{v_{j}^{2}}{(\varepsilon_{d}-\varepsilon_{j})^{2}}$,
using the results of \cite{A}:
\begin{equation}
f(\Gamma)=\gamma_{0}^{\frac{1}{2}}\Gamma^{-\frac{3}{2}}\exp\left(-\pi\gamma_{0}/\Gamma\right)
\label{eq:fGamma1}
\end{equation}
where the characteristic decay rate $\gamma_{0}$ is
\begin{equation}
\gamma_{0}=\frac{2\eta}{\hbar}\left[\frac{a_{0}n\langle
v(\textbf{r})\rangle}{W}\right]^{2} \label{eq:gamma0}
\end{equation}
$W$ is the width of energy fluctuation for the traps. $n$ is the
density of traps, while $a_{0} \sim O(1)$ is a constant with a
dimension of volume. Note that in the limit $\eta\rightarrow0$,
then $\gamma_{0}\rightarrow0$ and the probability distribution
function of decay rates is only nonzero when $\Gamma=0$, i. e. the
QD/SM electron is localized. Physically, the smallness of $\eta$
has to be compared with the energy difference
$\epsilon_{d}-\epsilon_{j}$ \cite{AGD}. As discussed in the
delocalized regime, the coupling of trap electrons with phonons
and non-radiative processes etc. can lead to a sufficiently small,
but non-zero value of $\eta$. The above results are applicable to
many topologically disordered systems \cite{LW}. Thus, for a QD/SM
in this regime, the single emitter experiences very long bright
periods, which are characterized by $1/\gamma_{0}$. The
\emph{on}-time probability distribution function is obtained from
(\ref{eq:fGamma1}):
\begin{equation}
P_{\text{on}}(t)=\left(\frac{t}{\pi\gamma_{0}}\right)^{-\frac{1}{2}}\exp\left(-\sqrt{4\pi\gamma_{0}t}\right)
\label{eq:Pon1}
\end{equation}
So in the limit $t\ll 1/\gamma_{0}$, the probability distribution
of the \emph{on}-time follows a power law with an exponent of
$-1/2$. On a much longer time scale, $P_{\text{on}}(t)$ deviates
from the power law and shows a stretched-exponential tail. This
stretched-exponential tail is of purely quantum nature and depends
on the degree of disorder $W/\langle v(\text{r})\rangle$.

Now we discuss the mechanism for the \emph{off}-time probability
distribution. Suppose a charge is on the trap site $j$. The self-energy
of $G_{j}(\omega)$ is $G_{d}(\omega)$. The decay rate is
vanishingly small, so the charge is localized. Then
\emph{on}-state can only be recovered via quantum tunnelling of
trapped electrons. $P_{\text{off}}(t)$ is therefore given by
\begin{equation}
P_{\text{off}}(t)=\int_{0}^{\infty}dr
g(r)\gamma_{\text{off}}(r)\exp\left[-\gamma_{\text{off}}(r)t\right]
\end{equation}
where $g(r)$ is the probability density of an excited electron
having been trapped at a distance of $r$ from the QD/SM and it is
proportional to the square of the excited state wave function.
$\gamma_{\text{off}}(r)$ is the decay rate of a trapped electron
to the QD/SM. A model based on this picture was first proposed by
Verberk \emph{et.al.} \cite{VOO} and experimentally examined in
\cite{IBC}. Since the typical size of a QD/SM is of the order
$30$\AA~in radius, so to a good approximation, the QD/SM can be
viewed as a shallow impurity \cite{K} and one can use effective
mass theory. The wave function of an excited QD/SM electron
behaves as $\exp(-r/a)$ at large distances, where
$a=\hbar^2\epsilon_{0}\kappa/m^{\ast}_{d}e^{2}=\tilde{a}\kappa$ is
the effective Bohr radius and is related to the ionization energy,
$E_{\text{ion}}$, of the excited electron by
$a=\hbar/\sqrt{2m^{\ast}_{d}E_{\text{ion}}}$. $\kappa$ is the
relative static dielectric constants of the media. $m^{\ast}_{d}$
is the effective mass of the QD/SM electron. Similarly,
$\gamma_{\text{off}}(r)$ is given by \cite{MA}:
$\gamma_{\text{off}}\exp(-2r/b)$, where
$b=\hbar/\sqrt{2m^{\ast}_{t}E_{t}}$ is the spatial extent of a
trapped electron's wave function with $m^{\ast}_{t}$ and $E_{t}$
being the effective mass and ionization energy at the trap site.
We obtain $P_{\text{off}}(t)\approx Ct^{-m_{\text{off}}}$ with
$C=a\Gamma(1+b/a)/\gamma_{\text{off}}^{b/a}b$ and
\begin{equation}
m_{\text{off}}=1+\frac{b}{a}=1+\sqrt{\frac{m^{\ast}_{d}E_{\text{ion}}}
{m^{\ast}_{t}E_{t}}}=1+\frac{b}{\tilde{a}}\frac{1}{\kappa}
\label{eq:moff}
\end{equation}
We see that the difference in $m_{\text{off}}$ for different
systems comes mainly from the effective mass $m^{\ast}_{d}$ and
the ability of a matrix to stabilize the charged QD/SM and the
ejected electron. In contrast to the diffusion model \cite{TM1},
the exponents of $P_{\text{off}}(t)$ depend on the relative static
dielectric constant $\kappa$ instead of the Cole-Davison $\beta$
parameter. This is consistent with recent experimental results
\cite{SCB1,IBC}. In addition, our model allows $m_{\text{off}}$ to
take a different value from $m_{\text{on}}$
\cite{KFHGN,CMB,SCB1,SCB2}, while the diffusion model predicts
that they must be the same.

To conclude, we have proposed a mechanism for FI as being a
manifestation of Anderson localization.  The \emph{on}-time
manifold is shown to be generated by different realizations of
electron delocalization from the QD/SM through the mechanism of
random resonance. The quantum theory predicts a universal
probability distribution function for the \emph{on} time and shows
$m_{\text{on}}=-2$ is indeed a robust result, which corresponds to
the long time limit of $P_{\text{on}}(t)$. The \emph{off} state
corresponds to a localized electron in the environment. The
recovery of the \emph{on} state is realized via quantum tunneling.

This work is supported by the National Science Foundation (NSF)
under the Grant No. CHE0306287.


\begin{thebibliography}{99}
\bibitem{NOTE1} The dressed electronic state is a linear
superposition of
$|0_{\text{g}},1_{\text{ex}},0_{\text{photon}}\rangle$ and
$|1_{\text{g}},0_{\text{ex}},1_{\text{photon}}\rangle$.
\bibitem{NOTE2} In the case when translational symmetry is
preserved, this corresponds to the conventional treatment of
changing summation to an integration weighted by energy density of
states and thus gives finite $\Gamma$.
\end{thebibliography}
\end{document}